\documentclass[twocolumn]{aastex62.mod}

\received{December 24, 2018}
\revised{February 5, 2019}
\accepted{February 6, 2019}

\shorttitle{The \textit{Hubble Space Telescope} UV Legacy Survey of
  Galactic Globular Clusters. XVIII}

\shortauthors{Libralato et al.}

\usepackage{xspace}
\usepackage{amsmath}
\usepackage{rotating}

\newcommand{\hst}{\textit{HST}\xspace}

\newcommand{\magv}{$m_{\rm F606W}$\xspace}
\newcommand{\magi}{$m_{\rm F814W}$\xspace}
\newcommand{\colvi}{$(m_{\rm F606W}-m_{\rm F814W})$\xspace}
\newcommand{\eqmaguv}{m_{\rm F275W}}
\newcommand{\eqmagu}{m_{\rm F336W}}
\newcommand{\eqmagb}{m_{\rm F438W}}
\newcommand{\eqmagv}{m_{\rm F606W}}
\newcommand{\eqmagi}{m_{\rm F814W}}

\begin{document}

\title{The \textit{Hubble Space Telescope} UV Legacy Survey of
  Galactic Globular Clusters. XVIII. Proper-motion kinematics of multiple stellar populations in the core regions of NGC~6352.}

\correspondingauthor{Mattia Libralato}
\email{libra@stsci.edu}

\author[0000-0001-9673-7397]{Mattia Libralato}
\affil{Space Telescope Science Institute 3700 San Martin Drive, Baltimore, MD 21218, USA}

\author[0000-0003-3858-637X]{Andrea Bellini}
\affil{Space Telescope Science Institute 3700 San Martin Drive, Baltimore, MD 21218, USA}

\author[0000-0002-9937-6387]{Giampaolo Piotto}
\affil{Dipartimento di Fisica e Astronomia, Universit\`a di Padova, Vicolo dell'Osservatorio 3, Padova, I-35122, Italy}
\affil{INAF-Osservatorio Astronomico di Padova, Vicolo dell'Osservatorio 5, Padova, I-35122, Italy}

\author[0000-0003-1149-3659]{Domenico Nardiello}
\affil{Dipartimento di Fisica e Astronomia, Universit\`a di Padova, Vicolo dell'Osservatorio 3, Padova, I-35122, Italy}
\affil{INAF-Osservatorio Astronomico di Padova, Vicolo dell'Osservatorio 5, Padova, I-35122, Italy}

\author[0000-0001-7827-7825]{Roeland P. van der Marel}
\affil{Space Telescope Science Institute 3700 San Martin Drive, Baltimore, MD 21218, USA}
\affil{Center for Astrophysical Sciences, Department of Physics \& Astronomy, Johns Hopkins University, Baltimore, MD 21218, USA}

\author[0000-0003-2861-3995]{Jay Anderson}
\affil{Space Telescope Science Institute 3700 San Martin Drive, Baltimore, MD 21218, USA}

\author[0000-0003-4080-6466]{Luigi R. Bedin}
\affil{INAF-Osservatorio Astronomico di Padova, Vicolo dell'Osservatorio 5, Padova, I-35122, Italy}

\author[0000-0003-2742-6872]{Enrico Vesperini}
\affil{Department of Astronomy, Indiana University, Bloomington, IN 47405, USA}

\begin{abstract}

We present the analysis of the radial distributions and kinematic
properties of the multiple stellar populations (mPOPs) hosted in the
globular cluster (GC) NGC~6352 as part of the \textit{Hubble Space
  Telescope} (\hst) ``UV Legacy Survey of Galactic Globular Clusters''
program. NGC~6352 is one of the few GCs for which the mPOP tagging in
appropriate color-magnitude diagrams is clear in all evolutionary
sequences. We computed high-precision stellar proper motions for the
stars from the cluster's core out to 75 arcsec ($\sim$1.5 core radii,
or $\sim$0.6 half-light radii). We find that, in the region explored,
first- and second-generation stars share the same radial distribution
and kinematic properties. Velocity dispersions, anisotropy radial
profiles, differential rotation, and level of energy equipartition,
all suggest that NGC~6352 is probably in an advanced evolutionary
stage, and any possible difference in the structural and kinematic
properties of its mPOPs have been erased by dynamical processes in the
core of the cluster. We also provide an estimate of the mass of blue
stragglers and of main-sequence binaries through kinematics alone. In
general, in order to build a complete dynamical picture of this and
other GCs, it will be essential to extend the analyses presented in
this paper to the GCs' outer regions where some memories of the
initial differences in the mPOP properties, and those imprinted by
dynamical processes, might still be present.

\end{abstract}


\section{Introduction}

\begin{table*}[th!]
  \caption{List of observations of NGC~6352 used in this paper.}
  \centering
  \label{tab:log}
  \begin{tabular}{cccccl}
    \hline
    \hline
    GO & PI & Instrument/Camera & Filter & $N$ $\times$ Exp. Time & Epoch \\
    \hline
    10775 & Sarajedini & ACS/WFC & F606W & $4 \times 140$ s , $1 \times 7$ s & 2006 April \\
    & & & F814W & $4 \times 150$ s , $1 \times 7$ s & \\
    12746 & Kong & ACS/WFC & F625W & $2 \times 150$ s & 2012 February \\
    & & & F658N & $1 \times 643$ s , $1 \times 650$ s & \\
    & & WFC3/UVIS & F336W & $5 \times 400$ s , $1 \times 410$ s & \\
    13297 & Piotto & WFC3/UVIS & F275W & $2 \times 706$ s & 2013 August \\
    & & & & $2 \times 800$ s & 2014 May \\
    & & & F336W & $2 \times 311$ s & 2013 August \\
    & & & & $2 \times 311$ s & 2014 May \\
    & & & F438W & $1 \times 58$ s & 2013 August \\
    & & & & $1 \times 72$ s & 2014 May \\
    \hline
  \end{tabular}
\end{table*}

The \textit{Hubble Space Telescope} (\hst) ``UV Legacy Survey of
Galactic Globular Clusters'' program \citep[GO-13297, PI: Piotto;
  see][]{2015AJ....149...91P} has started a systematic, photometric
analysis of the multiple stellar populations (mPOPs) hosted in
globular clusters (GCs). We now know that essentially all GCs studied
with the proper tools show the presence of mPOPs, and that the mPOPs
found in different GCs are characterized by a wide variety of
properties.

After the initial studies focused on the identification of mPOPs, we
worked on the full characterization of the mPOP properties. We have
analyzed most of the photometric pieces of information contained in
this Treasury survey, but the wealth of information contained in this
data set is still far from being completely explored and revealed. We
have begun to study the kinematic properties of the mPOPs thanks to
state-of-the-art proper motions (PMs) in light of the recent work on
this research field
\citep{2010ApJ...710.1032A,2013ApJ...771L..15R,2015ApJ...810L..13B,2018ApJ...853...86B,2018ApJ...861...99L,2018MNRAS.479.5005M}.

In this work, we continue the investigation of the GC NGC~6352 started
by \citet{2015MNRAS.451..312N}. NGC~6352 is a metal-rich ($\rm [Fe/H]
= -0.67$, \citealt{1997A&AS..121...95C}) GC in the Bulge direction
with a mass of $6.1 \times 10^4$ $M_\odot$
\citep{2019MNRAS.482.5138B}. \citet{2015MNRAS.451..312N} demonstrated
the presence of two populations clearly distinguishable in the
red-giant branch (RGB), sub-giant branch (SGB), main sequence (MS),
asymptotic giant branch (AGB), and horizontal branch (HB) with
UV-optical color-magnitude diagrams (CMDs). The first-generation (1G)
population (hereafter, POPa) has a primordial chemical composition,
while the second-generation (2G) population (POPb) is almost coeval
($\Delta$Age = $10\pm120$ Myr), slightly enhanced in He ($\Delta Y =
0.029\pm 0.006$), and in Na \citep[e.g.,][]{2009A&A...493..913F}.

In this paper, we focus our attention on the structural and kinematic
properties of the mPOPs in this cluster. We first compute \hst-based,
high-precision PMs to select members of NGC~6352 among the multitude
of (Bulge and Disk) field stars. We then separate POPa and POPb, and
study their radial distributions and internal kinematics.

\section{Data sets and reduction}\label{obs}

We made use of \texttt{\_flc} exposures\footnote{\texttt{\_flt} images
  pipeline corrected for the charge-transfer-efficiency defects as
  described in \citet{2010PASP..122.1035A}.} taken with the Wide-Field
Channel (WFC) of the Advanced Camera for Survey (ACS) and with the
Ultraviolet-VISible (UVIS) channel of the Wide-Field Camera 3
(WFC3). The complete list of observations\footnote{DOI reference:
  \protect\dataset[10.17909/t9-m7g1-8a93]{http://dx.doi.org/10.17909/t9-m7g1-8a93}} is
shown in Table~\ref{tab:log}.

\begin{figure*}
  \centering
  \includegraphics[width=1.\textwidth]{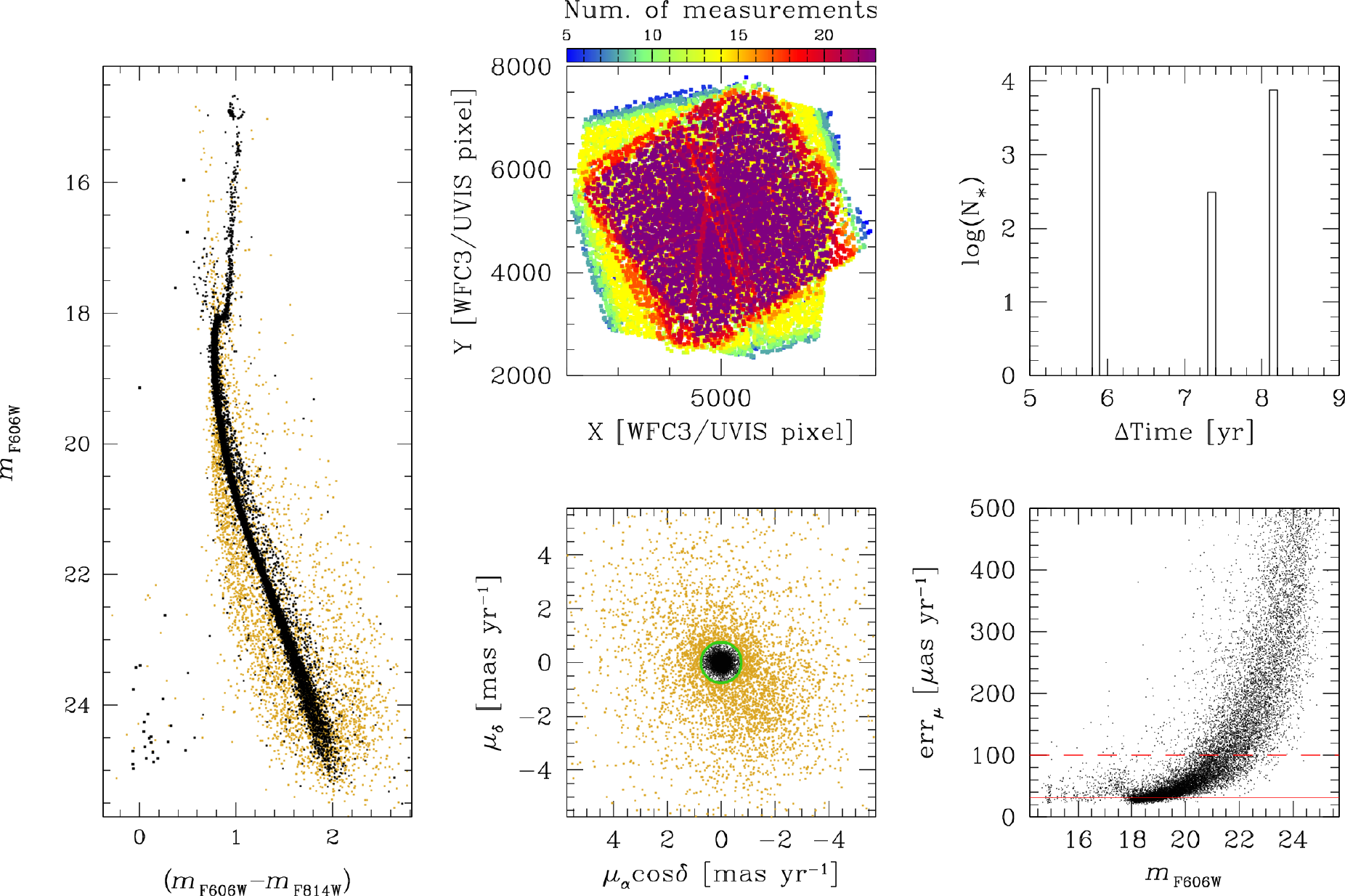}
  \caption{Overview of the PM catalog of NGC~6352. The \magv versus
    $(\eqmagv-\eqmagi)$ CMD (corrected for differential reddening, see
    Sect.~\ref{tag}) is presented in the left panel. Cluster stars
    (defined as those with a PM within 0.75 mas yr$^{-1}$ from the
    bulk motion of the cluster, i.e., $\sim$6 times the central
    velocity dispersion) are shown in black, while field objects are
    plotted in yellow. In the bottom-middle panel, we show the
    vector-point diagram of the stars in the FoV. By construction,
    cluster stars have a null bulk PM. NGC~6352 members are those
    within the green circle of radius 0.75 mas yr$^{-1}$. The
    top-middle panel shows the FoV covered by the observations. The
    cluster is centered at (5000,5000) WFC3/UVIS pixels. The center of
    the cluster is that defined by \citet{2010AJ....140.1830G}. Stars
    are color-coded as in the bar on top according to the number of
    measurements used in the PM fit. The logarithm of the number of
    sources as a function of the temporal baseline is shown in the
    top-right box. Finally, in the bottom-right panel we show the 1D
    PM error (err$_{\rm PM}$) in $\mu$as yr$^{-1}$ as a function of
    \magv. The solid red line is set at the median value of err$_{\rm
      PM} \sim 32$ $\mu$as yr$^{-1}$ for the best-measured stars
    ($18<\eqmagv<19$), while the dashed red line is set at 0.1 mas
    yr$^{-1}$ as reference.}
  \label{fig:overview1}
\end{figure*}

The data reduction is a combination of a first- and a second-pass
photometric stages, and was performed as in
\citet{2018ApJ...853...86B} and \citet{2018ApJ...861...99L}. We refer
to these papers for an extensive description of the work flow.

The first-pass photometry allows us to detect the brightest, most
isolated sources in each exposure in a single finding wave, and
measure position and flux of these objects via point-spread-function
(PSF) fit. The publicly-available, spatially-variable \hst library
PSFs\footnote{\href{http://www.stsci.edu/~jayander/STDPSFs/}{http://www.stsci.edu/$\sim$jayander/STDPSFs/}.}
are tailored to each exposure, and stellar positions are corrected for
geometric distortion using the state-of-the-art corrections available
for \hst detectors
\citep{2006acs..rept....1A,2009PASP..121.1419B,2011PASP..123..622B}. Positions
and fluxes are then used to build a common reference-frame system.

The second-pass photometry employs all images at once to enhance the
contribution of faint sources. All close-by neighbors are subtracted
from each image prior to estimate position and flux of a given source,
thus improving the measurements in crowded regions.

The main differences with \citet{2018ApJ...861...99L} are that (i) we
used the Gaia Data Release 2
\citep[DR2,][]{2016A&A...595A...1G,2018A&A...616A...1G} to set up
orientation (X and Y axes toward West and North, respectively) and
pixel scale (40 mas yr$^{-1}$) of our reference frame system, and (ii)
we run the second-pass photometry tool separately for the three data
sets (GO-10775, GO-12746, and GO-13297) as done in
\citet{2018ApJ...853...86B}, so as to measure stars that might have
moved by more than 1 pixel from one epoch to another.

We calibrated our photometry on to the Vega-mag system following the
prescriptions given in, e.g., \citet{2017ApJ...842....6B} and
\citet{2018MNRAS.481.3382N}. We measured bright, isolated stars on the
\texttt{\_drc} exposures using aperture photometry with aperture of 5
pixels, and corrected for the finite aperture. We then computed the
2.5$\sigma$-clipped median value of the magnitude difference between
our photometry and the aperture-corrected \texttt{\_drc}-based
magnitudes. Finally, we calibrated our photometry by adding to our
instrumental magnitudes this median difference and the Vega-mag
zero-point given in the STScI
website\footnote{\href{http://www.stsci.edu/hst/acs/analysis/zeropoints}{http://www.stsci.edu/hst/acs/analysis/zeropoints}
  for ACS/WFC and
  \href{http://www.stsci.edu/hst/wfc3/phot_zp_lbn}{http://www.stsci.edu/hst/wfc3/phot\_zp\_lbn}
  for WFC3/UVIS.}.

\begin{figure*}
  \centering
  \includegraphics[width=1.\textwidth]{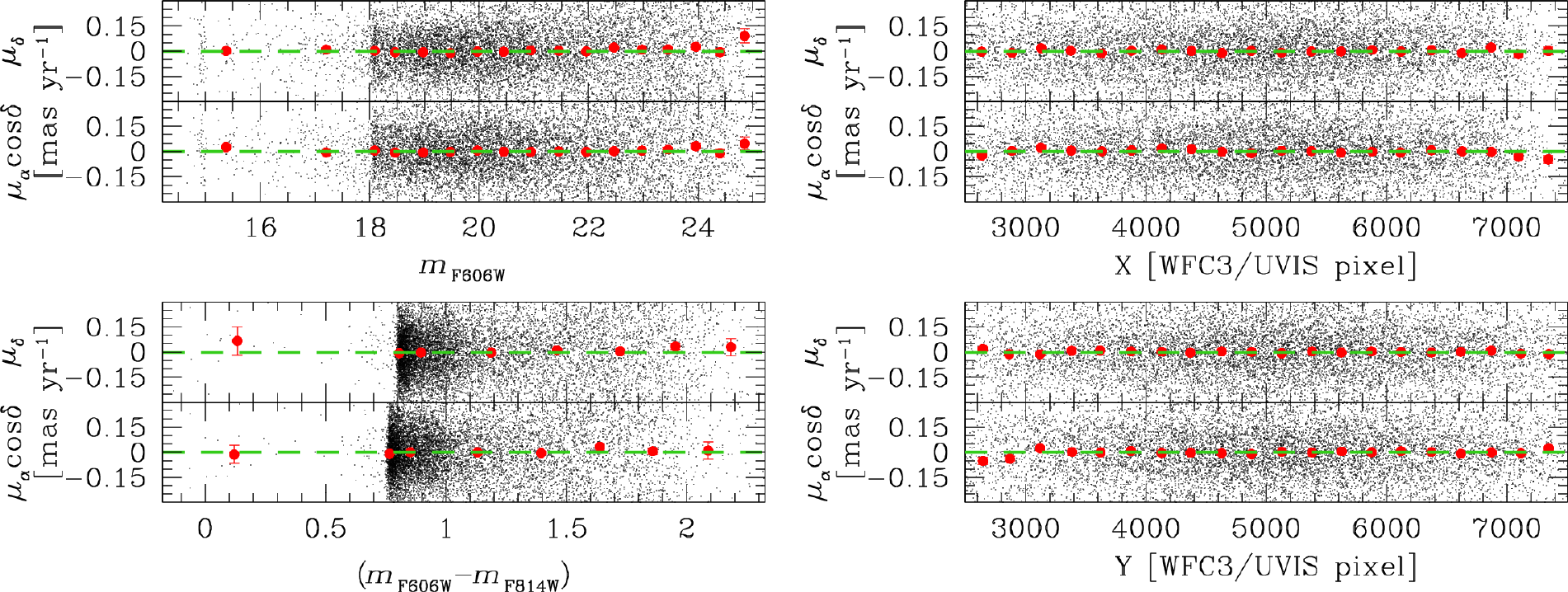}
  \caption{$\mu_\alpha \cos\delta$ and $\mu_\delta$ PMs as a function
    of the \magv, $(\eqmagv-\eqmagi)$, X and Y master-frame
    positions. Only cluster stars with a PM lower than 0.75 mas
    yr$^{-1}$ are shown. The red points (with error bars) are the
    3.5$\sigma$-clipped median values of the PMs in different
    magnitude, color or position bins. The green, dashed lines are set
    at $\mu_\alpha \cos\delta$ and $\mu_\delta$ equal to 0.}
  \label{fig:overview2}
\end{figure*}

PMs were computed by combining the multi-epoch \hst data as in
\citet{2014ApJ...797..115B}, to which we refer for the detailed
description of the method. We also corrected for spatially-variable,
high- and low-frequency systematic effects following the descriptions
of \citet{2014ApJ...797..115B,2018ApJ...853...86B} and
\citet{2018ApJ...861...99L}. An overview of the final PM catalog of
NGC~6352 is presented in Fig.~\ref{fig:overview1}. The median 1D PM
error of stars at the level of the MS turn-off ($18 < \eqmagv < 19$)
is of $\sim$32 $\mu$as yr$^{-1}$. For comparison, stars in this field
at the same magnitude level have a median 1D PM error of $\sim$360
$\mu$as yr$^{-1}$ in the Gaia DR2. Figure~\ref{fig:overview2} shows
the PM along $\alpha \cos\delta$ and $\delta$ directions as a function
of the \magv magnitude, $(\eqmagv-\eqmagi)$ color, X and Y
master-frame positions. No clear systematic trends at levels
comparable to the random errors for the best-measured stars arise from
these plots.

Only objects measured in all the available filters of GO-10775 and
GO-13297 were considered in the analysis, thus the field of view (FoV)
at disposal is that covered by the GO-13297 WFC3/UVIS data. Stars are
considered as well measured if: (i) their quality of PSF
(\texttt{QFIT}) parameter is larger than the 95-th percentile at any
given magnitude\footnote{The closer the \texttt{QFIT} is to 1, the
  better is the PSF fit. In addition to the 95-th percentile cut, we
  consider as well measured all objects with a \texttt{QFIT} value
  higher than 0.99. We also discard all sources with \texttt{QFIT}
  lower than 0.6.}, (ii) their magnitude rms is lower than the 95-th
percentile at any given magnitude\footnote{Similarly as for the
  \texttt{QFIT}, we also keep all sources with a magnitude rms lower
  than 0.05 mag, and discard those with a rms higher than 0.4 mag.},
(iii) they are measured in at least 40\% of the images, (iv) their
fraction of neighbor flux within the fitting radius before neighbor
subtraction is less than 1, (v) their shape parameter \texttt{RADXS}
\citep[excess/deficiency of flux outside of the fitting radius with
  respect to the PSF prediction, see][]{2008ApJ...678.1279B} is lower
than the 85-th percentile at any given magnitude, (vi) their flux is
at least 3$\sigma$ above the local sky, (vii) their reduced $\chi^2$
of the PM fit is lower than 2, and (viii) their rejection rate in the
PM fit is lower than 30\%.

Finally, we also excluded stars with a PM error larger than half the
local velocity dispersion $\sigma_\mu$ of the closest 100 cluster
stars, and stars outside a radius of 75 arcsec from the cluster center
(i.e., the four corners of the FoV) to avoid edge effects. These last
two conditions are applied only in the kinematic analysis.

\section{Dissecting NGC~6352}\label{tag}

We identified the two populations of NGC~6352 in all evolutionary
sequences by combining UV and optical CMDs and color-color diagrams,
similarly to what is described in \citet{2017MNRAS.464.3636M} and
\citet{2018ApJ...861...99L}.

We initially corrected for the differential reddening affecting our
photometry following the prescription of, e.g., \citet{2012A&A...540A..16M} and
\citet{2017ApJ...842....7B}. NGC~6352 is in the direction of the Bulge
and the extinction is high \citep[$E(B-V)=0.22$,][2010
  edition]{1996AJ....112.1487H}.  We assumed that cluster members are
affected by the same amount of intra-cluster differential
reddening. For each star, we selected a sample of close-by cluster
stars and measured their median shift along the reddening direction
with respect to a fiducial line in the CMD. Since NGC~6352 hosts two
distinct populations, we selected the most populous group (POPb) to
define the fiducial line. We tailored the correction for each CMD we
present in this paper.

\begin{figure*}
  \centering
  \includegraphics[width=1.\textwidth]{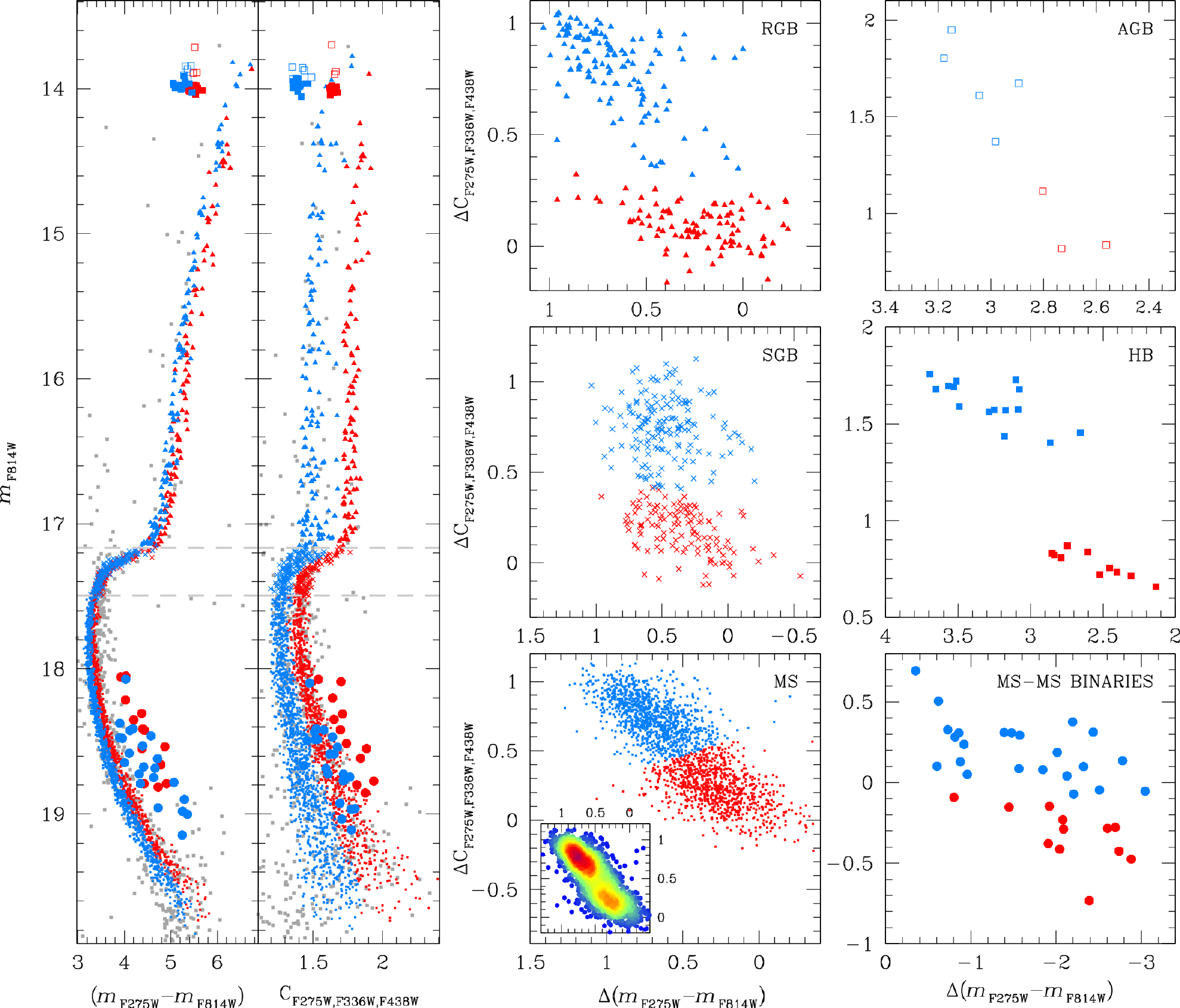}
  \caption{This figure illustrates the mPOP tagging performed to
    disentangle POPa (red points) and POPb (azure points). We
    rectified the \magi versus $(\eqmaguv-\eqmagi)$ and \magi versus
    C$_{\rm F275W,F336W,F438W}$ CMDs (left panels). The gray, dashed
    horizontal lines in the left CMDs set the RGB, SGB, and MS
    separations. Gray dots are well-measured stars that are not
    selected as POPa,b stars or equal-mass binaries. We then computed
    the chromosome maps of each evolutionary sequence and separated 1G
    and 2G stars. For the MS stars, we made use of the Hess diagram in
    the inset of the MS chromosome map at this end. Equal-mass MS
    binaries of POPa and POPb were defined as described in the text
    and in Fig.~\ref{fig:overviewbin}.}
  \label{fig:overviewphot}
\end{figure*}

We first defined the RGB, SGB, and MS regions, and then separated POPa
and POPb by means of the pseudo two-color diagrams ``chromosome maps''
\citep{2017MNRAS.464.3636M}. We drew by hand two fiducial lines along
RGB, SGB and MS in the \magi versus $(\eqmaguv-\eqmagi)$ and \magi
versus C$_{\rm
  F275W,F336W,F438W}$$=$$(\eqmaguv-\eqmagu)-(\eqmagu-\eqmagb)$ planes,
and rectified these sequences by defining $\Delta$color$=$$(\rm
color-fiducial_{\rm red})/(fiducial_{\rm blue}-fiducial_{\rm red})$,
where ``color'' is either $(\eqmaguv-\eqmagi)$ or C$_{\rm
  F275W,F336W,F438W}$.

The chromosome maps of NGC~6352 are presented in
Fig.~\ref{fig:overviewphot}. The primordial POPa and the 2G stars of
POPb are shown in red and azure, respectively. The two populations are
quite obvious to separate in the RGB, SGB, AGB, and HB. The
identification of the two populations in the MS is based on the Hess
diagram of the chromosome map (inset in the MS panel of
Fig.~\ref{fig:overviewphot}).

\begin{figure}
  \centering
  \includegraphics[width=1.\columnwidth]{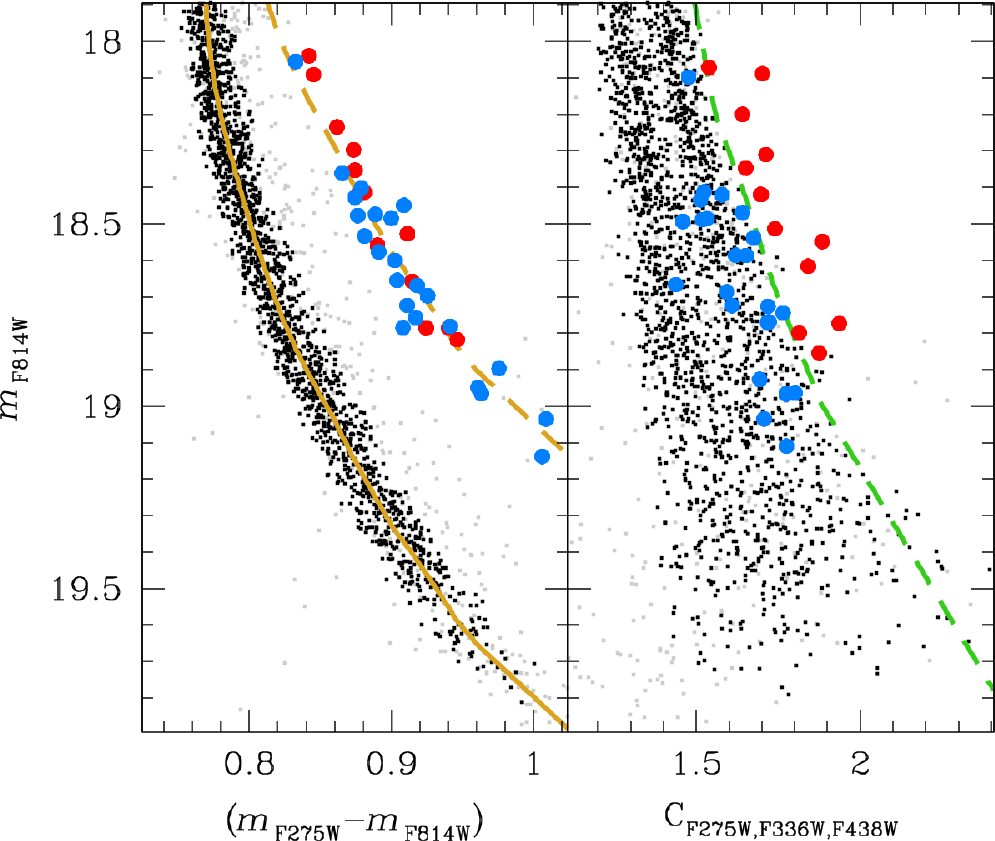}
  \caption{Overview of the mPOP tagging along the equal-mass MS
    binaries. In the \magi versus $(\eqmagv-\eqmagi)$ CMD (left panel)
    we initially selected as equal-mass MS binaries objects 0.75 mag
    brighter than the MS fiducial line (yellow, solid line). The MS
    fiducial line shifted by 0.75 mag is shown as a yellow, dashed
    line in the CMD as reference. Black points are stars belonging to
    either POPa or POPb, while gray dots are all other objects. We
    then defined POPa/ POPb equal-mass MS binaries stars
    brighter/fainter than the reddest edge (green, dashed line) of the
    POPa MS in the \magi versus C$_{\rm F275W,F336W,F438W}$ CMD (right
    panel).}
  \label{fig:overviewbin}
\end{figure}

We also separated equal-mass binaries that seem to belong to POPa and
POPb as follows. First, we defined equal-mass MS binaries objects
$\sim$0.75 mag brighter than the MS fiducial in the \magi versus
$(\eqmagv-\eqmagi)$ CMD (left panel of
Fig.~\ref{fig:overviewbin}). Then, in the \magi versus C$_{\rm
  F275W,F336W,F438W}$ CMD, we tagged as POPa binaries the equal-mass
binaries brighter than the reddest edge of the POPa MS. The remaining
binaries belong to the POPb (right panel of
Fig.~\ref{fig:overviewbin}). Note that this classification for the
equal-mass binaries does not consider the case of equal-mass mixed
1G$+$2G binaries, which are expected theoretically
\citep[e.g.,][]{2016MNRAS.457.4507H} and might have different
properties from 1G$+$1G and 2G$+$2G systems.

\section{mPOPs spatial distribution}\label{distr}

Our \hst observations of NGC~6352 cover a FoV out to $\sim$2$r_{\rm
  c}$ \citep[$r_{\rm c} = 49.8$ arcsec,][2010
  edition]{1996AJ....112.1487H} from the cluster center. We analyzed
the radial distributions of the populations along the RGB, SGB, and MS
as follows. We chose as workbench the \magi versus C$_{\rm
  F275W,F336W,F438W}$ CMD. We rectified the sequences by means of
fiducial lines as described in Sect.~\ref{tag}. The fiducial lines
were made by linearly interpolating the median color and magnitude in
different (0.5, 0.1, and 0.5 mag for RGB, SGB, and MS, respectively)
magnitude bins.

We computed the histogram of the $\Delta$C$_{\rm F275W,F336W,F438W}$
color adopting a bin width of 0.05 mag. This methodology is however
sensitive to the bin width and the starting point of the histogram. To
ensure a bias-free estimate of the fraction of POPa and POPb stars, we
computed the histograms 10\,000 times, each time adding a random noise
to the points, and averaged them. The noise added to each star was
randomly picked from a Gaussian distribution with $\sigma$ equal to
the C$_{\rm F275W,F336W,F438W}$ photometric error of the star. This
way we removed the dependencies on the bin width (by blurring or
sharpening the distributions) and on the starting point of the
histogram (by shifting the stars in the rectified CMD).

The histograms of RGB, SGB, and MS stars present two distinct peaks.
We fitted the final average histograms with a pair of Gaussian
functions and estimated the fraction of stars belonging to POPa and
POPb in a statistical fashion as described in
\citet{2013ApJ...765...32B}.

\begin{figure*}
  \centering
  \includegraphics[width=0.75\textwidth]{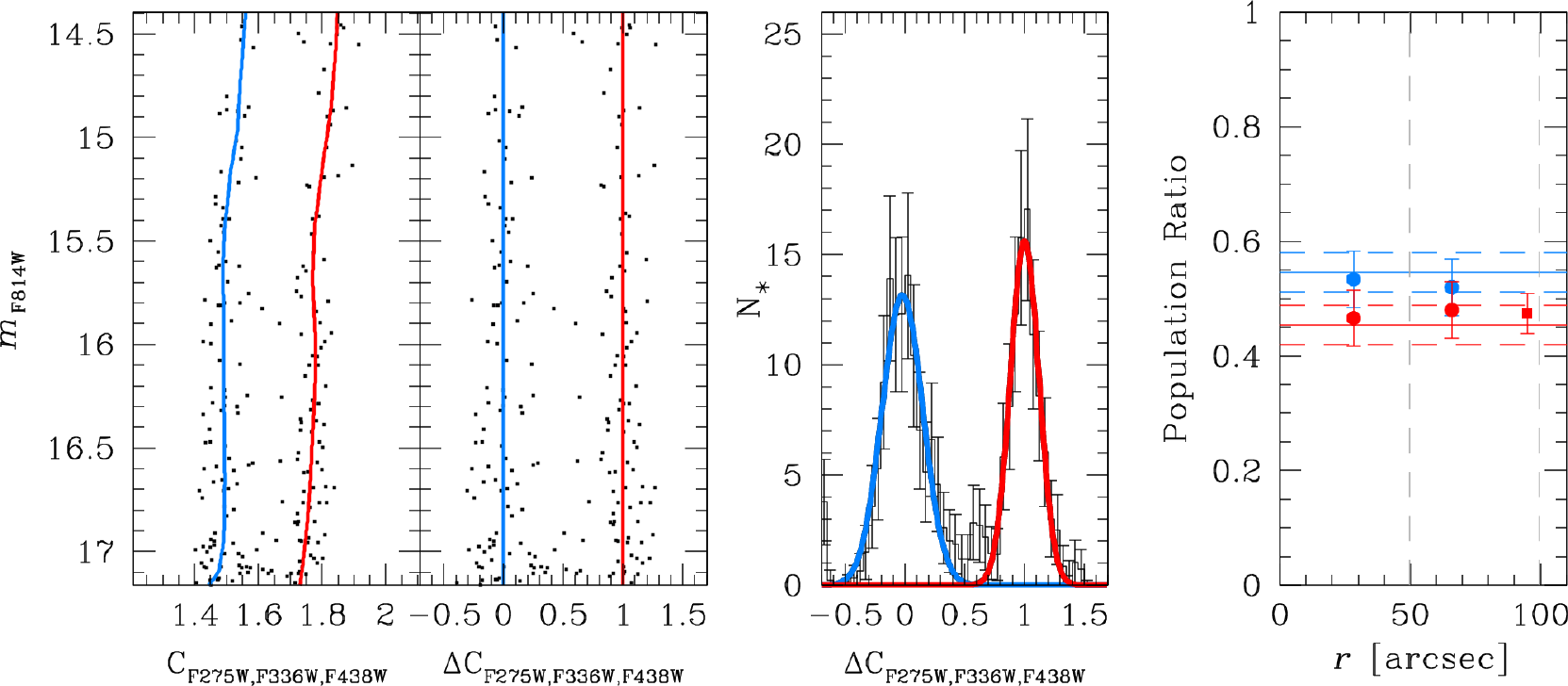}
  \caption{Left: \magi versus C$_{\rm F275W,F336W,F438W}$ CMD of the
    RGB stars over the entire FoV. The azure and red lines represent
    the fiducial lines used to rectify the CMD in this magnitude
    range. Middle-left: \magi versus $\Delta$C$_{\rm
      F275W,F336W,F438W}$ CMD. Middle-right: Dual-Gaussian fit of the
    $\Delta$C$_{\rm F275W,F336W,F438W}$ distribution. The azure and
    red lines represent the individual Gaussian fit (the dual Gaussian
    fit is not shown because the two distributions are too
    separated). The histogram is the average of 10\,000 histograms
    (see the text for details). Right: Radial trends of the population
    ratio of RGBa (red filled circles) and RGBb (azure filled
    circles). The solid horizontal lines represent the value of the
    ratios obtained by considering all stars in the FoV. The dashed
    lines are set at $\pm 1\sigma$ with respect to the solid
    lines. The two gray, dashed vertical lines are set at $r_{\rm c}$
    and $2r_{\rm c}$. Finally, the red square (with error bars) marks
    the POPa ratio computed by \citet{2017MNRAS.464.3636M} for the
    same cluster.}
  \label{fig:rgb_rd}
  \includegraphics[width=0.75\textwidth]{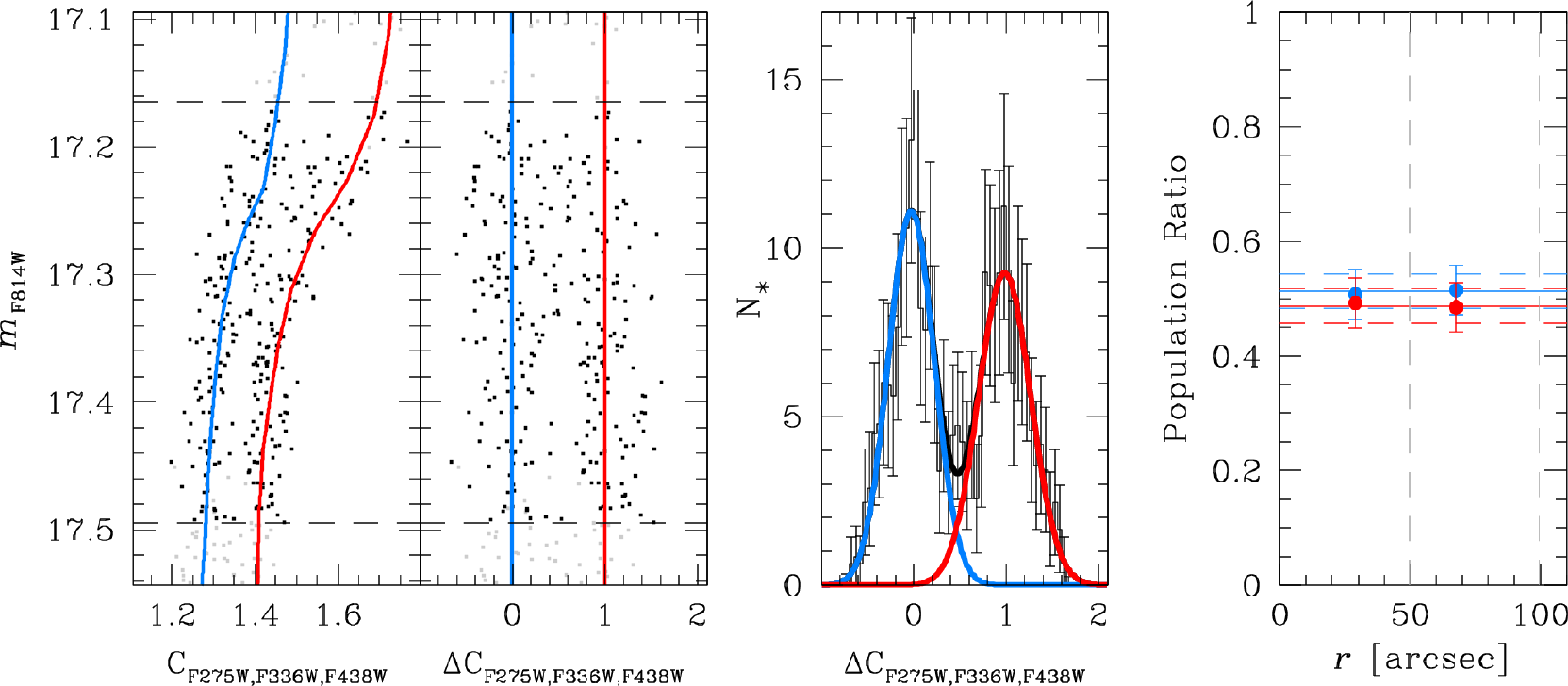}
  \caption{Similar to Fig.~\ref{fig:rgb_rd}, but for SGB stars. Stars
    not included in the SGB analysis (i.e., that are outside the
    magnitude interval defined by the two gray, dashed horizontal
    lines in the left and middle-left CMDs) are shown with gray
    dots. The black line in the middle-right panel represents the dual
    Gaussian fit.}
  \label{fig:sgb_rd}
  \includegraphics[width=0.75\textwidth]{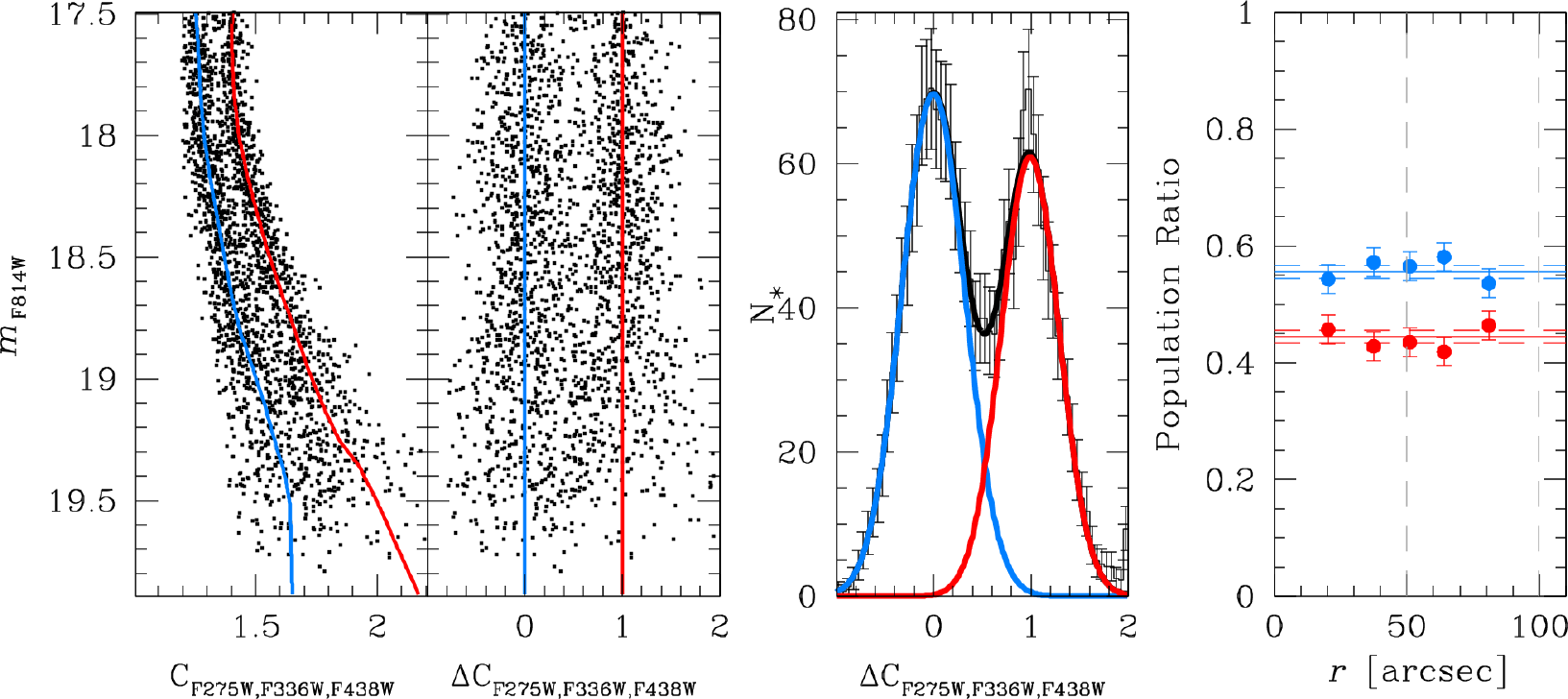}
  \caption{Similar to Figs.~\ref{fig:rgb_rd} and \ref{fig:sgb_rd}, but
    for MS stars.}
  \label{fig:ms_rd}
\end{figure*}

Figures~\ref{fig:rgb_rd}, \ref{fig:sgb_rd} and \ref{fig:ms_rd} present
the radial distributions of the mPOPs along the RGB, SGB, and MS,
respectively. The left panels show the \magi versus C$_{\rm
  F275W,F336W,F438W}$ CMD of the stars in the entire FoV. The
rectified CMDs are displayed in the middle-left panels. The fiducial
lines used to rectify the CMD are shown as azure and red lines. The
average histogram and the dual Gaussian functions for all stars in the
FoV are shown in the middle-right panels. The ratios
POPa,b/(POPa+POPb) over the entire FoV are shown as red and azure
solid horizontal lines, respectively, in the rightmost panels. The
dashed horizontal lines indicate the $\pm 1\sigma$ thresholds.

We repeated the same procedure shown in these three panels for stars
in equally-populated radial bins (two bins for the RGB, two for the
SGB and five for the MS). The population ratio computed in each radial
bin is shown with a filled circle in the rightmost panels of
Figs.~\ref{fig:rgb_rd}, \ref{fig:sgb_rd} and \ref{fig:ms_rd}. The
number of bins in the RGB and SGB was chosen to have about 100 stars
in each bin, while for the MS five bins are a good compromise between
the need of a high statistics in each bin (about 360 stars per bin)
and map possible radial features in the distributions. We also changed
the number of radial bins for the MS and found consistent results
within the error bars.

The ratios POPa/(POPa+POPb) are listed in Table~\ref{tab:raddist}. We
find that the population ratio does not vary with the distance from
the cluster center, and POPb is slightly more numerous than POPa. Our
analysis is focused on the innermost regions of the cluster, which are
the first parts where any initial radial gradient in the population
ratio is erased by the effects of dynamical evolution, as shown in
\citet{2013MNRAS.429.1913V}. Our result is therefore consistent with
the theoretical expectations. Considering that NGC~6352 has a
relatively short half-mass relaxation time \citep[t$_{\rm hm}\simeq
  0.8$ Gyr,][2010 edition]{1996AJ....112.1487H}, it is quite possible
that the two populations are completely mixed over the entire cluster
extension, but data covering a larger radial interval are necessary to
further explore this issue.

Finally, it is worth noticing that \citet{2017MNRAS.464.3636M}
estimated the global fraction of 1G RGB stars in NGC~6352. They found
a ratio of POPa RGB stars equal to $0.474 \pm 0.035$, in agreement
with our independent estimate for the RGBa stars ($0.452 \pm 0.035$).

\paragraph{HB, AGB, and equal-mass MS binaries}

The number of stars in the AGB and HB is too small to perform an
analysis of their radial distributions. However, the relative number
of POPa stars is $\sim$38\% and 40\% of the total for AGB and HB,
respectively, in agreement with the results for RGB, SGB, and MS. We
also find the relative number of POPa stars for the equal-mass MS
binaries is $\sim$33\%.

\begin{figure}
  \centering
  \includegraphics[width=\columnwidth]{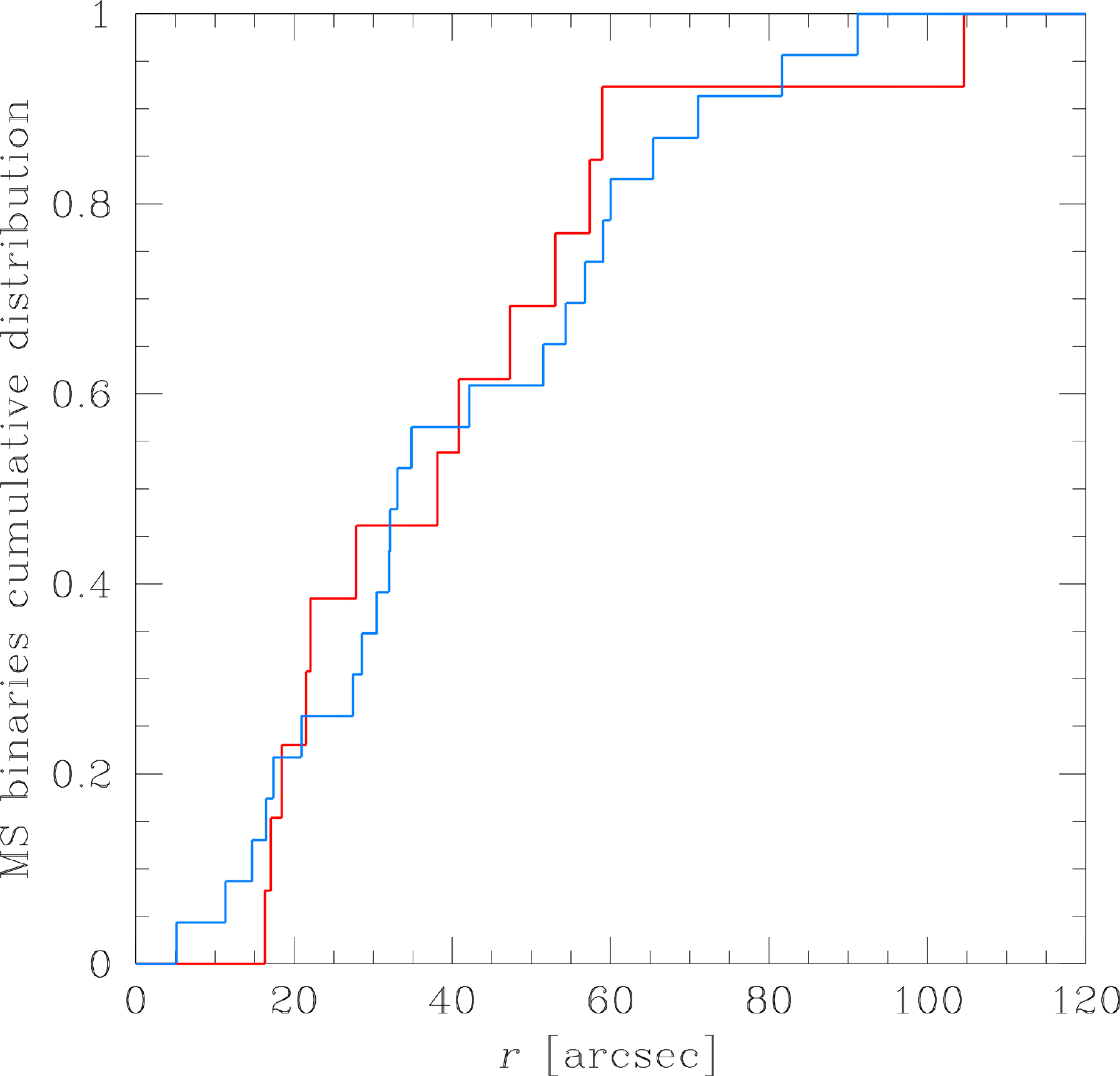}
  \caption{Cumulative distributions of equal-mass 1G (red) and 2G
    (azure) MS binaries.}
  \label{fig:bin_rd}
\end{figure}

In Fig.~\ref{fig:bin_rd}, we present the cumulative radial
distribution of equal-mass binaries of POPa and POPb. The binary stars
of the two populations are mixed in the cluster innermost
regions. Recently, \citet{2018MNRAS.tmp.3147H} have shown that
binary-star spatial mixing can be delayed by a number of dynamical
processes affecting binaries (ionization, recoil, ejection). A more
extended radial coverage than the one available in our \hst data is
required to test whether binaries are not mixed yet or NGC~6352 is
sufficiently dynamically old to have reached complete spatial mixing
also for its binary stars.

\begin{table}
  \caption{Results of the radial-distribution analysis for NGC~6352.}
  \centering
  \label{tab:raddist}
  \begin{tabular}{cccc}
    \hline
    \hline
    $r_{\rm min}$ & $r_{\rm max}$ & $\overline{r}$ & POPa/(POPa+POPb) \\
    $[$arcsec$]$ & $[$arcsec$]$ & $[$arcsec$]$ & \\
    \hline
    \multicolumn{4}{c}{\textbf{RGB}} \\
    \hline
      0.0 & 93.80 & 47.14 & $0.452 \pm 0.035$ \\
    \hline
      0.0 & 46.09 & 28.31 & $0.475 \pm 0.050$ \\
    46.09 & 93.80 & 65.97 & $0.485 \pm 0.050$ \\
    \hline
    \multicolumn{4}{c}{\textbf{SGB}} \\
    \hline
      0.0 & 101.66 & 48.37 & $0.487 \pm 0.030$\\
    \hline
      0.0 &  47.87 & 28.95 & $0.492 \pm 0.044$\\
    47.87 & 101.66 & 67.65 & $0.485 \pm 0.044$\\
    \hline
    \multicolumn{4}{c}{\textbf{MS}} \\
    \hline
      0.0 & 103.29 & 50.77 & $0.445 \pm 0.011$\\
    \hline
      0.0 &  30.34 & 20.34 & $0.457 \pm 0.025$\\
    30.34 &  44.11 & 37.49 & $0.428 \pm 0.024$\\
    44.11 &  57.89 & 51.04 & $0.435 \pm 0.025$\\
    57.89 &  70.55 & 63.98 & $0.419 \pm 0.024$\\
    70.55 & 103.29 & 80.93 & $0.464 \pm 0.025$\\
    \hline
  \end{tabular}
\end{table}

\section{Internal kinematics of NGC~6352}\label{kin}

We analyzed the internal kinematics of the members of NGC~6352. In the
following analysis, the velocity dispersions were obtained by
correcting the observed scatter of the PMs for the uncertainties of
the individual PMs as in \citet{2010ApJ...710.1063V}. Velocity
dispersions in mas yr$^{-1}$ were converted in km s$^{-1}$ by assuming
a cluster distance of 5.6 kpc \citep[2010
  edition]{1996AJ....112.1487H}.

\subsection{mPOP kinematics}\label{mpopkin}

\begin{figure*}
  \centering
  \includegraphics[width=\textwidth]{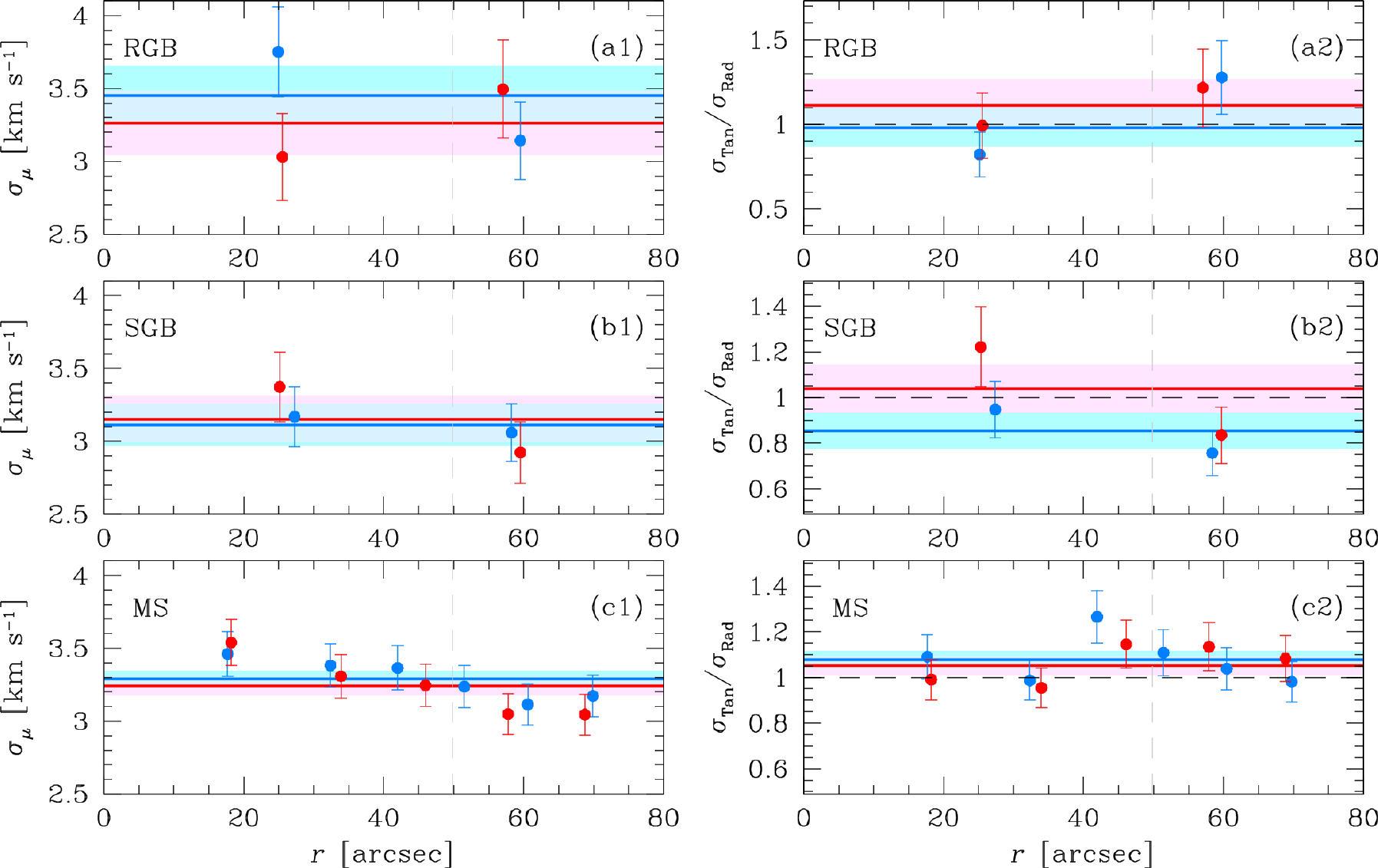}
  \caption{Velocity-dispersion (left) and anisotropy radial profiles
    (right) of RGB (top panels), SGB (middle panels), and MS (bottom
    panels) stars of NGC~6352. Red (azure) points represent the
    average values of POPa (POPb) stars in each radial bin. The core
    radius is indicated by the gray dashed vertical lines. The
    horizontal lines are set at the average value of all POPa and POPb
    stars; the pink and cyan shaded areas mark the corresponding
    $\pm$1$\sigma$ regions. The black, dashed lines in the right
    panels are set at 1 as reference.}
  \label{fig:mpopall}
\end{figure*}

We analyzed the combined ($\sigma_\mu$), radial ($\sigma_{\rm Rad}$),
and tangential ($\sigma_{\rm Tan}$) velocity dispersions as a function
of distance from the cluster center for each mPOP separately and in
equally-populated radial bins. Figure~\ref{fig:mpopall} presents the
velocity-dispersion (left) and anisotropy radial profiles (right) for
the mPOPs along the RGB (top), SGB (middle), and MS (bottom),
respectively. The value of each point is computed using 32 (43) stars
for RGBa (RGBb), 55 (65) stars for SGBa (SGBb), 145 (150) stars for
MSa (MSb), respectively.  As a reference, we also draw the average
velocity dispersion and anisotropy of the entire sample (horizontal
lines) and the corresponding $\pm$1$\sigma$ uncertainties (shaded
regions).

Figure~\ref{fig:mpopall} shows that POPa (1G) and POPb (2G) have the
same kinematics and are kinematically isotropic within
$\sim$2$\sigma$. Considering that the data probe the cluster's
innermost regions within $\sim$1.5$r_{\rm c}$, or $\sim$0.6$r_{\rm h}$
\citep[$r_{\rm c} = 49.8$ and $r_{\rm h} = 123.6$ arcsec,][2010
  edition]{1996AJ....112.1487H}, this result is expected. The
innermost regions are the first to relax and no significant anisotropy
is, in general, expected close to the cluster's center \citep[see,
  e.g.,][]{2016MNRAS.455.3693T}. For example,
\citet{2013ApJ...771L..15R} and \citet{2015ApJ...810L..13B} found the
presence of some velocity anisotropy in the 2G stars at distances
larger than 1.5-2 $r_{\rm h}$. A study of the kinematics in the
cluster's outermost regions would be necessary to further explore the
possible presence of anisotropy in the velocity distribution due to
internal dynamical processes and the cluster's interaction with the
external tidal field \citep[see, e.g.,][]{2016MNRAS.455.3693T}.

\paragraph{HB and equal-mass MS binaries}

\begin{figure}
  \centering
  \includegraphics[width=\columnwidth]{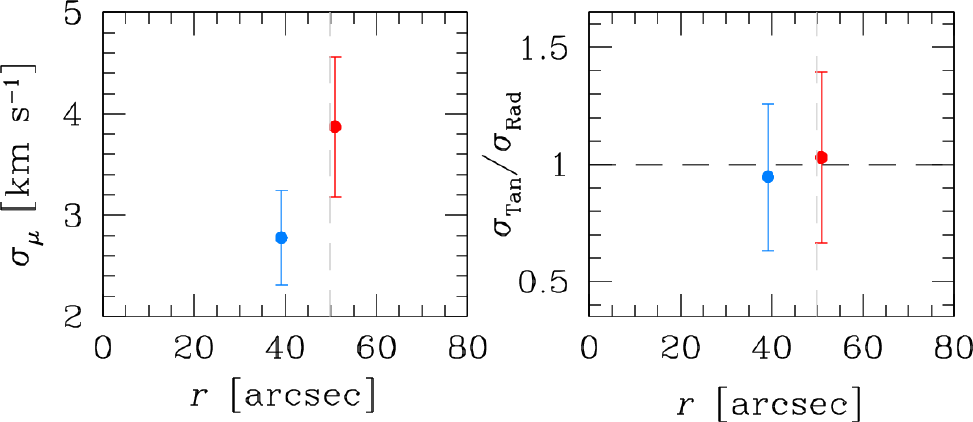}
  \caption{Combined velocity dispersion (left panel) and anisotropy
    (right panel) as a function of radius for the two populations in
    the HB of NGC~6352. Each bin is made up by 10 HB stars.}
  \label{fig:mpophb}
\end{figure}

We analyzed the internal kinematics of HB stars as a
whole. Figure~\ref{fig:mpophb} shows $\sigma_\mu$ and the anisotropy
for the HB stars (red and azure points for POPa and POPb,
respectively) as a function of radius. Again, the two populations
hosted along the HB of NGC~6352 are isotropic and share the same
kinematics within the errors.

\begin{figure}
  \centering
  \includegraphics[width=\columnwidth]{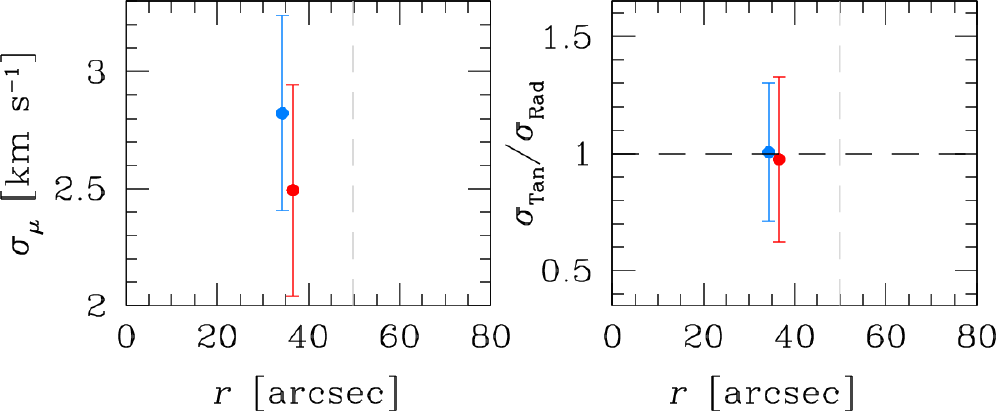}
  \caption{$\sigma_\mu$ as a function of distance from the cluster
    center for equal-mass MS binaries is shown in the left panel. Each
    bin represents the average value of all binaries (11 and 16 for
    POPa and POPb, respectively). The right panel shows the
    tangential-to-radial anisotropy profile for the same stars.}
  \label{fig:mpopbin}
\end{figure}

We also computed the velocity dispersions of the mPOPs along the
MS-binary sequence. Due to the low-number statistics, we made only one
radial bin containing all stars in the FoV. The results are presented
in Fig.~\ref{fig:mpopbin}. The azure and red points show the
velocity-dispersion of MS binaries. MS binaries of POPa and POPb share
the same kinematics and are isotropic.

We were not able to analyze AGB stars because of the very few stars at
disposal.

\subsection{mPOP differential rotation}

\begin{figure}
  \centering
  \includegraphics[width=\columnwidth]{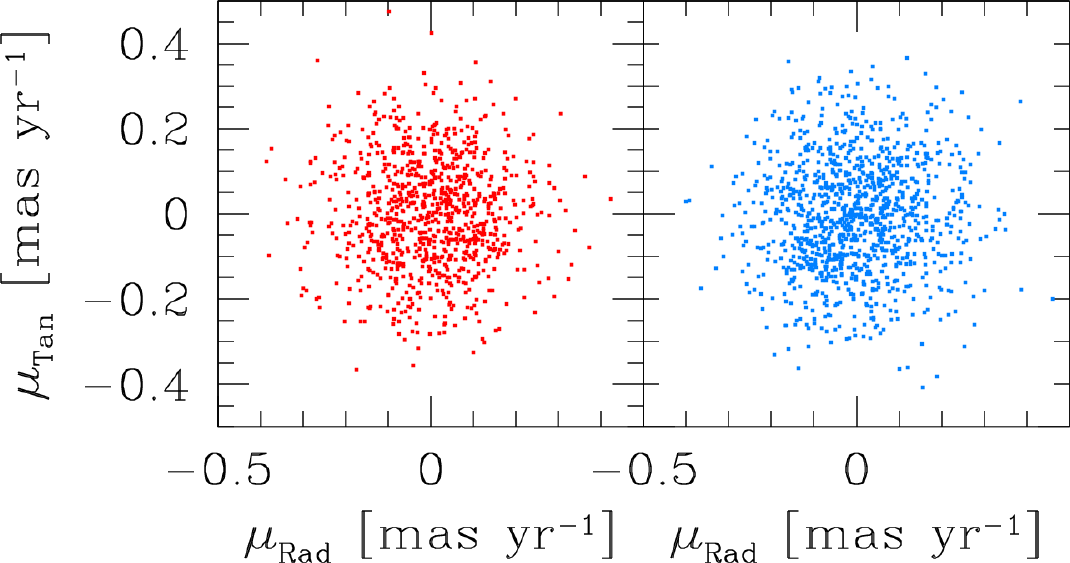}
  \caption{POPa (left panel) and POPb (right panel) PMs in the
    $\mu_{\rm Tan}$ versus $\mu_{\rm Rad}$ plane.}
  \label{fig:mpoprot}
\end{figure}

We investigated the possible presence of differential rotation. We
cannot directly measure any signature of cluster's rotation in the
plane of the sky because the systemic rotation signal is absorbed by
the six-parameter linear transformations. However, there are different
methods to infer the presence of rotation that rely on the skewness of
the PMs in the tangential direction
\citep[see][]{2017ApJ...850..186H,2018ApJ...853...86B,2018ApJ...861...99L}. Figure~\ref{fig:mpoprot}
shows the PMs of NGC~6352 in the $\mu_{\rm Tan}$ versus $\mu_{\rm
  Rad}$ plane.

We measured the amount of skew in the $\mu_{\rm Tan}$ with (1) the
value $G_1$ and the significance test $Z_{G1}$ \citep{Cramer1997}, and
(2) the third-order Gauss-Hermite moment $h_3$
\citep{1993ApJ...407..525V}. We find
\begin{equation}
  \textrm{POPa:}\left\{
  \begin{array}{l}
    G_1 = 0.09 \textrm{ , } Z_{G_1} = 1.05 \\
    h_3 = 0.021 \pm 0.026 \\
  \end{array}
  \right. \, ,
\end{equation}
\begin{equation}
  \textrm{POPb:}\left\{
  \begin{array}{l}
    G_1 = 0.09 \textrm{ , } Z_{G_1} = 1.27 \\
    h_3 = 0.021 \pm 0.024 \\
  \end{array}
  \right. \, . \\
\end{equation}
These values are consistent with an absence of differential rotation
in NGC~6352.

Recently, \citet{2018MNRAS.481.2125B} estimated the amount of rotation
in GCs by means of the Gaia DR2 PMs. The authors measured for NGC~6352
$\mu_{\rm Tan}(r<3r_{\rm h}) = 0.001^{+0.014}_{-0.013}$ mas yr$^{-1}$
and a 1$\sigma$ upper limit of $V/\sigma = 0.09$ ($V$ is the value of
the rotation peak, $\sigma$ is the central velocity dispersion):
values consistent with no rotation in the plane of the
sky. Furthermore, they found a correlation between $V/\sigma$ and the
relaxation time.

As GCs advance in their evolution, they gradually lose their initial
rotation \citep[see, e.g.,][and references
  therein]{2007MNRAS.377..465E,2017MNRAS.469..683T}. The absence of
rotation in NGC~6352 could be another indication of its advanced
dynamical age.

\subsection{Level of Energy Equipartition}\label{equip}

Globular clusters evolve toward a state of increasing energy
equipartition ($\sigma_\mu \propto m^{-\eta}$ with $m$ the stellar mass
and $\eta=0.5$ for a complete energy equipartition), but they are not
expected to reach a complete equipartition
\citep{2013MNRAS.435.3272T,2016MNRAS.458.3644B,2017MNRAS.464.1977W}.

We divided the MS in 7 equally-populated magnitude bins, and computed
$\sigma_\mu$ and median magnitude of the $\sim$230 stars in each
bin. We then transformed these magnitudes in stellar masses by using a
set of Dartmouth isochrones \citep{2008ApJS..178...89D}.

NGC~6352 has an age of $\sim$13.0 Gyr
\citep[e.g.,][]{2017MNRAS.468.1038W}, $\rm [Fe/H] = -0.67$
\citep{1997A&AS..121...95C}, $\rm [\alpha/Fe] = 0.4$, $E(B-V) = 0.22$,
and it is at a distance of 5.6 kpc \citep[2010
  edition]{1996AJ....112.1487H}. \citet{2015MNRAS.451..312N} analyzed
the He content and relative age of the two populations hosted in
NGC~6352, finding $\Delta Y = 0.029 \pm 0.006$ and $\rm \Delta Age =
10 \pm 120$ Myr.

\begin{figure*}
  \centering
  \includegraphics[width=\textwidth]{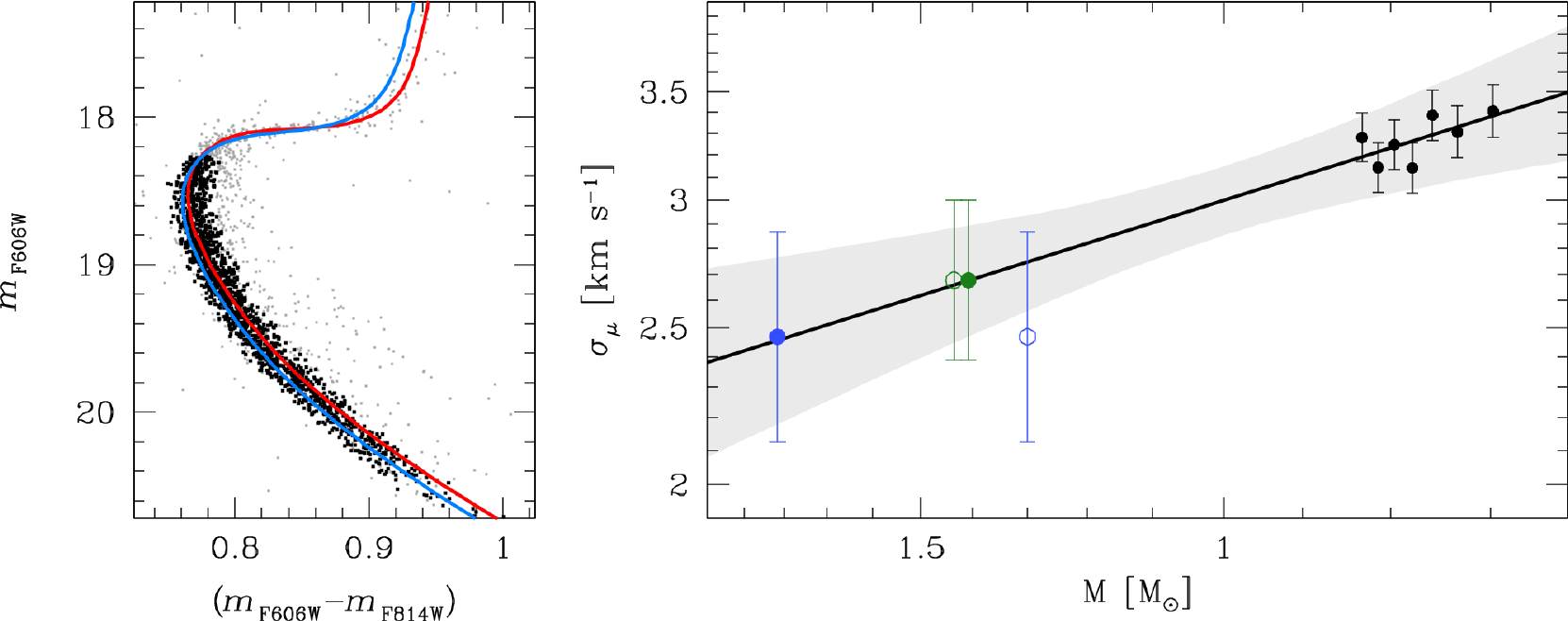}
  \caption{The \magv versus \colvi CMD of NGC~6352 is presented in the
    left panel. Black points represent stars used to measure the level
    of energy equipartition, gray dots are all other stars. The red
    and azure lines are the best-fit isochrones of the primordial and
    He-enhanced populations, respectively. In the right panel, we show
    $\sigma_\mu$ versus mass in the log-log plane. The black dots
    correspond to the MS stars highlighted in black in the CMD. The
    black line is the best fit to these points. The slope of such line
    gives a direct estimate of the level of energy equipartition
    $\eta$. The gray area represents the 1$\sigma$ confidence region
    of the straight-line fit. The filled blue and green dots are the
    values of BSs and equal-mass MS binaries, respectively. The filled
    dots were obtained by assuming that BSs and binaries have reached
    the same energy-equipartition state of the MS stars, thus lying on
    the black straight line in the plot. We used this assumption to
    measure the mass of these objects (Sect.~\ref{mass}). The open
    dots are instead obtained by assuming the theoretical value of the
    mass of these objects. These open points lie on the straight line
    within 1$\sigma$, meaning that BS and equal-mass MS binaries have
    the same level of energy equipartition of lower-mass members of
    NGC~6352.}
  \label{fig:equip}
\end{figure*}

The best-fit isochrones are derived using a technique similar to that
described in \citet{2015MNRAS.451..312N} to estimate the He difference
between POPa and POPb in NGC~6352. We compared the colors of the
observed fiducial lines of the two populations in the \magv versus
$(\eqmagv-\eqmagi)$ CMD with the colors of the fiducial lines of
synthetic CMDs built from a set of isochrones with $\rm [Fe/H] =
-0.67$ and $\rm [\alpha/Fe] = 0.4$. We treated both mPOPs at the same
time, using the same isochrones except for the He content ($Y=0.256$
for POPa and $Y=0.285$ for POPb). We let distance, reddening and age
vary, and defined the best-fit isochrones as those that provided the
lowest $\chi^2$. We find
\begin{equation}
  \left\{
  \begin{array}{rl}
    E(B-V) & = 0.25 \\
    \textrm{Distance} & = 5.3 \textrm{ kpc} \\
    \textrm{Age} & = 13.0 \textrm{ Gyr} \\
  \end{array}
  \right.
\end{equation}
The best isochrones for POPa and POPb are shown in the left panel of
Fig.~\ref{fig:equip} as red and azure lines, respectively. The MS of
NGC~6352 is made up by 45\% of POPa stars and 55\% of POPb
stars. Therefore, we weighted the median mass in each magnitude bin by
these ratios.

Finally, we fitted the $\sigma_\mu$ versus mass values in a log-log
plane with a weighted least-square straight line and find
\begin{equation}
  \eta = 0.334 \pm 0.182 \phantom{,}
\end{equation}
The result is presented in the right panel of
Fig.~\ref{fig:equip}. The black points picture the velocity
dispersions of the seven MS bins; the black line is the best fit to
these points [We postpone the explanation of the azure and green
  points to the next section.]. We computed the value of $\eta$ by
using isochrones with slightly different ages, $\rm [Fe/H]$, and by
neglecting the presence of the mPOPs, and obtained consistent results
within the errors.

This value of $\eta$ provides further evidence that NGC~6352 is in an
advanced stage of its dynamical evolution, and it is in general
consistent with the theoretical predictions of
\citet{2013MNRAS.435.3272T} and \citet{2017MNRAS.464.1977W}, although
our uncertainty on the value of $\eta$, mainly due to the low
statistics and the small mass range considered, is large.

\subsection{The mass of blue stragglers and MS binaries}\label{mass}

Blue stragglers (BSs) and MS-binary systems are more massive than a
typical (single) star in a GC. If the GC has some degree of energy
equipartition, we expect these massive objects to have a lower
velocity dispersion than the other stars. \citet{2016ApJ...827...12B}
computed the mass of the BSs in several GCs by assuming that the
velocity dispersions and masses of these stars scale as
\begin{equation}
  \alpha = \frac{\sigma_{\rm BSS}}{\sigma_{\rm MSTO}} =
  \left(\frac{M_{\rm BSS}}{M_{\rm MSTO}}\right)^{-\eta} \phantom{,}.
\end{equation}
We calculated the mass of the BSs in NGC~6352 in a similar fashion,
and extended the same methodology to the equal-mass MS binaries.

\begin{figure*}
  \centering
  \includegraphics[width=\textwidth]{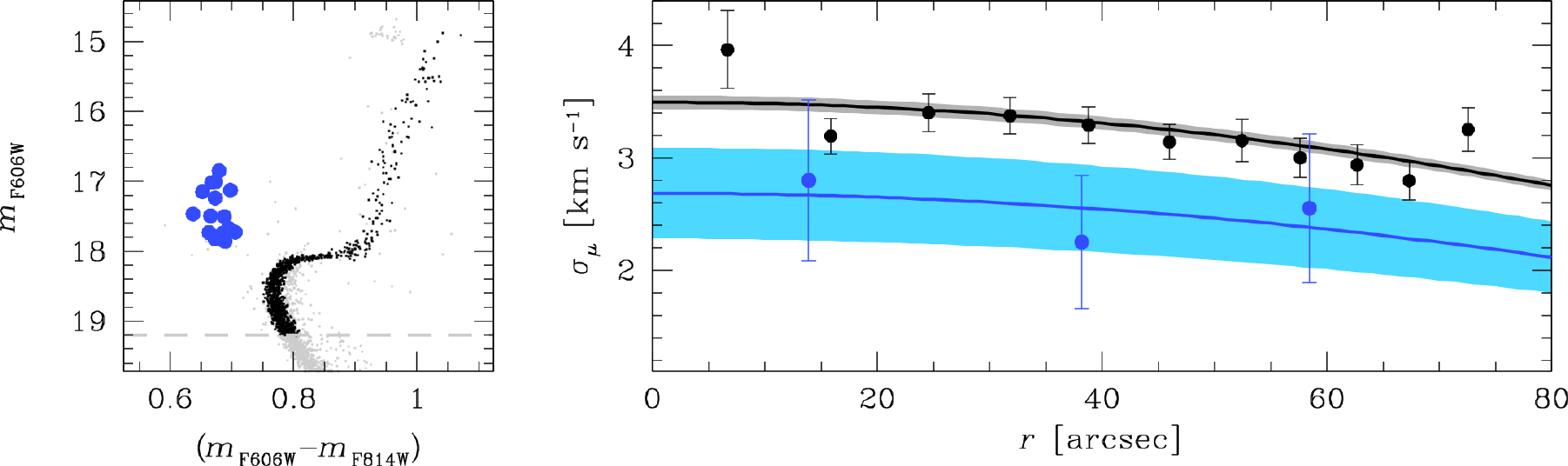}
  \caption{Highlight of the procedure adopted to measure the mass of
    the BSs. We compared the combined velocity dispersion of the BSs
    (blue dots in the \magv versus \colvi CMD in the left panel) with
    that of all stars brighter than 0.5 mag below the MS turn-off
    (black dots brighter than the gray, dashed line in the CMD; other
    objects are shown as gray dots). The right panel shows the
    velocity dispersion of BSs (blue dots) and of the reference
    population (black dots). It is clear that massive objects move
    slower than lighter stars because of the energy equipartition. We
    fit the velocity-dispersion radial profile of the reference
    population (black line), and then found the value $\alpha$ that
    provides the best fit of the same profile on the BS kinematics
    (blue line). The gray and pale-blue regions set the 1$\sigma$
    errors of the fitted profiles for reference stars and BSs,
    respectively.}
  \label{fig:bss}
\end{figure*}

For the BSs (blue dots in the left panel of Fig.~\ref{fig:bss}), we
computed the combined velocity dispersion of these stars in 3
equally-populated radial bins of 5 stars each. We chose as reference
population all RGB, SGB, and MS stars brighter than 0.5 mag below the
MS turn-off (black points in the CMD in Fig.~\ref{fig:bss}), and
computed the velocity dispersions of these reference objects in 11
radial bins\footnote{One bin of 37 stars was defined by all stars
  within 10 arcsec from the center of the GC. Five equally-populated
  bins of 118 stars each were defined from 10 arcsec to the core
  radius, and five bins of 81 stars each from the core radius to the
  limit of the analyzed FoV.}. The velocity-dispersion radial profiles
of BSs and reference stars are shown in the right panel of
Fig.~\ref{fig:bss} in blue and black, respectively.

For each radial bin of the reference population, we drew 10\,000
realizations of the average velocity dispersion by adding a random
Gaussian noise with $\sigma$ equal to the error in the velocity
dispersion. We then interpolated the median values of these
distributions by means of a 3rd-order polynomial function with a flat
center. The best fit is shown in Fig.~\ref{fig:bss}. The gray region
represents the 1$\sigma$ error of the fitted profiles.

We performed a similar computation for the BS, but this time we
interpolated the velocity-dispersion profile with the same 3rd-order
polynomial function of the reference population rescaled by a factor
$\alpha$. The blue line in Fig.~\ref{fig:bss} depicts the best fit for
the BSs, the pale-blue region is the 1$\sigma$-error region.

We find $\alpha = 0.77 \pm 0.12$. Given a MS turn-off mass of 0.83
$M_\odot$ and assuming that the SGB and RGB stars have the same
dynamical mass of those at the MS turn-off, we derive a mass for the
BSs of $1.82 \pm 0.37$ $M_\odot$. Our estimate is slightly larger,
although in agreement at 1$\sigma$ level, than the typical mass of a
BS \citep[between 1.0 and 1.6 $M_\odot$; see,
  e.g.,][]{2018ApJ...860...36F}.

\begin{figure*}
  \centering
  \includegraphics[width=\textwidth]{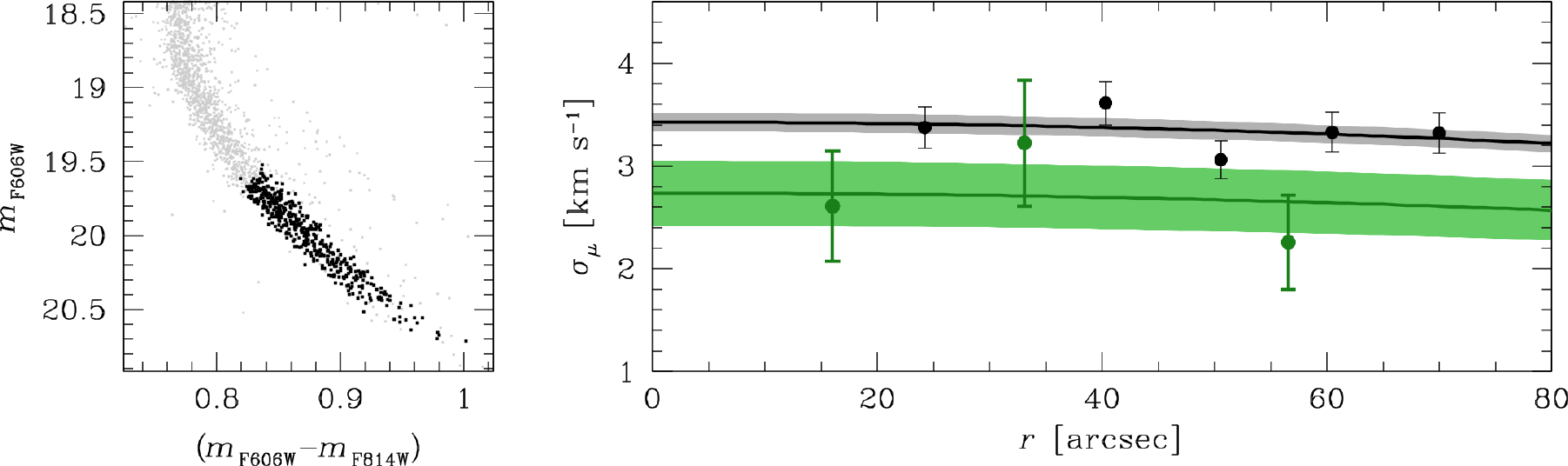}
  \caption{Similar to Fig.~\ref{fig:bss}, but for the equal-mass MS
    binaries.}
  \label{fig:msbin}
\end{figure*}

We also estimated the average mass of the equal-mass MS binary systems
(green dots in Fig.~\ref{fig:msbin}) as done for the BSs. This time we
chose as reference the MS stars 0.75 mag fainter than the equal-mass
binary region (black points in Fig.~\ref{fig:msbin}). We find $\alpha
= 0.80 \pm 0.09$. The average mass of the MS stars (computed as
described in Sect.~\ref{equip}) in the magnitude interval considered
in the analysis is $\sim$0.72 $M_\odot$, which gives us an average
mass of the binary systems of $1.41 \pm 0.30$ $M_\odot$. This is
consistent with a system made by two stars of 0.72 $M_\odot$ each.

Returning back to Fig.~\ref{fig:equip}, its right panel shows the
average velocity dispersion of all BSs and MS binaries as a function
of mass. The filled dots refer to the mass computed in this Section
and, by construction, they are aligned with the best fit of the MS
stars from which we computed the value of $\eta$. If we assume an
average mass of the BSs and MS binaries of 1.3 $M_\odot$ \citep[the
  mid-range value of typical BS masses from][]{2018ApJ...860...36F}
and 1.44 $M_\odot$ (twice the average MS stars at the same color level
of the binaries), respectively, we can have an independent check of
the goodness of our estimate of $\eta$. The empty dots plotted by
assuming these theoretical values of the mass for BSs and MS-binary
systems seem to be in agreement with the prediction of the MS stars
(black line), and confirm the level of energy equipartition we
computed with this \hst data set.

\subsection{Comparison with the literature}

As an external check, we made a comparison with the work of
\citet{2019MNRAS.482.5138B}. The authors fitted $N$-body models to
ground-based radial velocities, \hst-based mass functions, and the
Gaia DR2 PMs, and derived structural parameters for 144 Galactic
GCs\footnote{\href{https://people.smp.uq.edu.au/HolgerBaumgardt/globular/}{https://people.smp.uq.edu.au/HolgerBaumgardt/globular/}.}.

\begin{figure}
  \centering
  \includegraphics[width=\columnwidth]{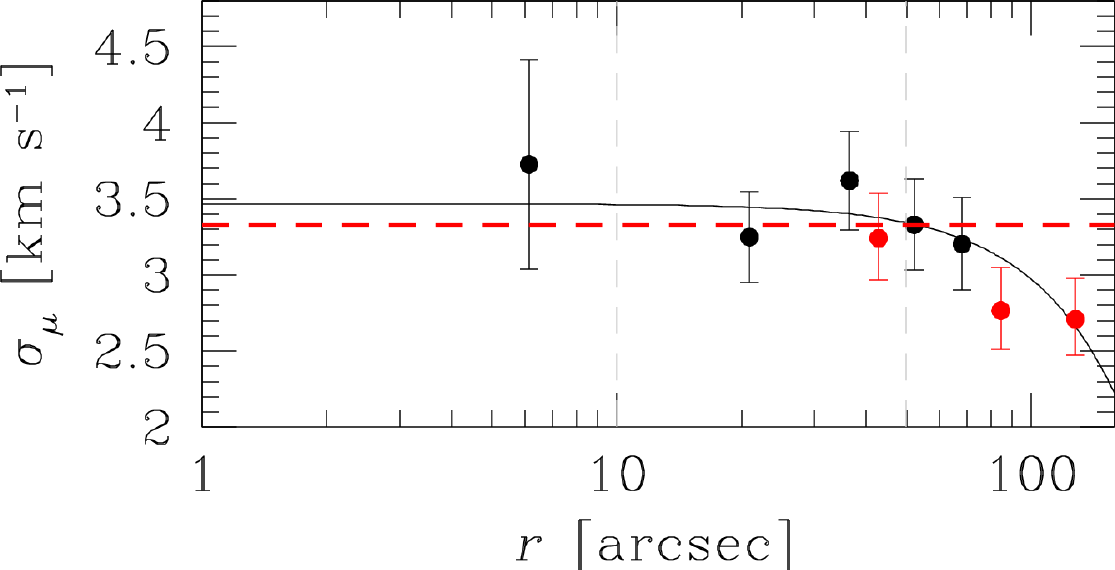}
  \caption{Combined velocity dispersion as a function of distance from
    the cluster center for RGB stars. Black points represent the
    kinematics of the RGB stars measured in this work, red points are
    those obtained by \citet{2019MNRAS.482.5138B} with the Gaia DR2
    PMs. The black line is the best fit to our data. The red, dashed
    line is set at the central value of the velocity dispersion
    computed by \citet{2019MNRAS.482.5138B}.}
  \label{fig:rgbprof}
\end{figure}

First, we compared our \hst-based velocity-dispersion radial profiles
with those obtained with the Gaia DR2. We limited our analysis to the
RGB stars to be consistent with \citet{2019MNRAS.482.5138B}. We
computed the average $\sigma_\mu$ of all stars within 10 arcsec from
the cluster center, and in 4 equally-populated (35 stars per bin)
radial bins in the remaining part of the FoV. The values of
$\sigma_\mu$ computed in this work are shown in black in
Fig.~\ref{fig:rgbprof}. The red points are the velocity dispersions
computed with the Gaia DR2 PMs by \citet{2019MNRAS.482.5138B},
rescaled to the cluster distance adopted in our work. The \hst- and
Gaia-based velocity dispersions are in excellent agreement with each
other.

\citet{2019MNRAS.482.5138B} provided an estimate of the central
velocity dispersion $\sigma_0 = 3.5$ km s$^{-1}$, or $\sigma_0 = 3.33$
km s$^{-1}$ upon rescaling for the distance adopted in our work. We
fitted our velocity-dispersion radial profiles with a 3rd-order
polynomial (forced to be flat at the center) and extrapolated a value
of $\sigma_0 = 3.47$ km s$^{-1}$, consistent with the value of
\citet{2019MNRAS.482.5138B}.

Finally, \citet{2019MNRAS.482.5138B} also computed the value of $\eta$
for stars in the mass range 0.5-0.8 $M_\odot$. They find $\eta =
0.36$, which is consistent with the value we derived from our
\hst-based PMs in Sect.~\ref{equip}.

\section{Conclusions}

In this paper we presented a comprehensive characterization of the
structural and kinematic properties of the mPOPs in the innermost
regions of the Galactic GC NGC~6352.

We used data from the \hst UV survey GO-13297 to identify 1G (POPa)
and 2G (POPb) stars along the RGB, SGB, and MS, finding a similar
fraction of POPa stars ($\sim$45\%) in each evolutionary group. We
also find no evidence of a significant variation in the fraction of
POPa stars with distance from the cluster center. As shown in previous
theoretical studies, the innermost regions are the first where any
initial radial gradients is erased, and the lack of a radial variation
in the fraction of stars belonging to the two populations is therefore
consistent with the theoretical expectations.

We studied the cluster's internal kinematics by means of \hst-based,
high-precision PMs. There is no sign of internal rotation, anisotropy
in the velocity distributions, and differences between the kinematic
properties of POPa and POPb stars. We also measured the dependence of
the velocity dispersion on the stellar mass, and provided a
quantitative estimate of the level of energy equipartition reached in
the cluster's inner regions.

NGC~6352 is probably in its advanced evolutionary stages, and any
difference in the structural and kinematic properties of mPOPs might
have been erased by dynamical processes over the entire cluster's
extension. It will be essential to extend the analysis presented in
this paper to the cluster' outer regions where some memories of the
initial or dynamically-induced differences might still be retained.

\section*{Acknowledgments}

M.L. and A.B. acknowledge support from STScI grant GO-13297. The
authors thank the anonymous Referee for the thoughtful suggestions
that improved the quality of the paper. Based on observations with the
NASA/ESA \textit{HST}, obtained at the Space Telescope Science
Institute, which is operated by AURA, Inc., under NASA contract NAS
5-26555. This work has made use of data from the European Space Agency
(ESA) mission {\it Gaia} (\url{https://www.cosmos.esa.int/gaia}),
processed by the {\it Gaia} Data Processing and Analysis Consortium
(DPAC,
\url{https://www.cosmos.esa.int/web/gaia/dpac/consortium}). Funding
for the DPAC has been provided by national institutions, in particular
the institutions participating in the {\it Gaia} Multilateral
Agreement.

\bibliographystyle{aasjournal}

\begin{thebibliography}{}

\bibitem[Anderson \& King(2006)]{2006acs..rept....1A} Anderson, J., \& King, I.~R.\ 2006, Instrument Science Report ACS 2006-01, 34 pages,

\bibitem[Anderson \& Bedin(2010)]{2010PASP..122.1035A} Anderson, J., \& Bedin, L.~R.\ 2010, \pasp, 122, 1035

\bibitem[Anderson \& van der Marel(2010)]{2010ApJ...710.1032A} Anderson, J., \& van der Marel, R.~P.\ 2010, \apj, 710, 1032

\bibitem[Baldwin et al.(2016)]{2016ApJ...827...12B} Baldwin, A.~T., Watkins, L.~L., van der Marel, R.~P., et al.\ 2016, \apj, 827, 12

\bibitem[Baumgardt et al.(2019)]{2019MNRAS.482.5138B} Baumgardt, H., Hilker, M., Sollima, A., \& Bellini, A.\ 2019, \mnras, 482, 5138 

\bibitem[Bedin et al.(2008)]{2008ApJ...678.1279B} Bedin, L.~R., King, I.~R., Anderson, J., et al.\ 2008, \apj, 678, 1279-1291

\bibitem[Bellini \& Bedin(2009)]{2009PASP..121.1419B} Bellini, A., \& Bedin, L.~R.\ 2009, \pasp, 121, 141

\bibitem[Bellini et al.(2011)]{2011PASP..123..622B} Bellini, A., Anderson, J., \& Bedin, L.~R.\ 2011, \pasp, 123, 622

\bibitem[Bellini et al.(2013)]{2013ApJ...765...32B} Bellini, A., Piotto, G., Milone, A.~P., et al.\ 2013, \apj, 765, 32

\bibitem[Bellini et al.(2014)]{2014ApJ...797..115B} Bellini, A., Anderson, J., van der Marel, R.~P., et al.\ 2014, \apj, 797, 115

\bibitem[Bellini et al.(2015)]{2015ApJ...810L..13B} Bellini, A., Vesperini, E., Piotto, G., et al.\ 2015, \apjl, 810, L13

\bibitem[Bellini et al.(2017a)]{2017ApJ...842....6B} Bellini, A., Anderson, J., Bedin, L.~R., et al.\ 2017a, \apj, 842, 6

\bibitem[Bellini et al.(2017b)]{2017ApJ...842....7B} Bellini, A., Anderson, J., van der Marel, R.~P., et al.\ 2017b, \apj, 842, 7

\bibitem[Bellini et al.(2018)]{2018ApJ...853...86B} Bellini, A., Libralato, M., Bedin, L.~R., et al.\ 2018, \apj, 853, 86 

\bibitem[Bianchini et al.(2016)]{2016MNRAS.458.3644B} Bianchini, P., van de Ven, G., Norris, M.~A., Schinnerer, E., \& Varri, A.~L.\ 2016, \mnras, 458, 3644

\bibitem[Bianchini et al.(2018)]{2018MNRAS.481.2125B} Bianchini, P., van der Marel, R.~P., del Pino, A., et al.\ 2018, \mnras, 481, 2125

\bibitem[Carretta \& Gratton(1997)]{1997A&AS..121...95C} Carretta, E., \& Gratton, R.~G.\ 1997, \aaps, 121, 95 

\bibitem[Cramer(1997)]{Cramer1997} Cramer, D., 1997, ``Basic Statistics for Social Research''. Routledge.

\bibitem[Dotter et al.(2008)]{2008ApJS..178...89D} Dotter, A., Chaboyer, B., Jevremovi{\'c}, D., et al.\ 2008, \apjs, 178, 89-101

\bibitem[Ernst et al.(2007)]{2007MNRAS.377..465E} Ernst, A., Glaschke, P., Fiestas, J., Just, A., \& Spurzem, R.\ 2007, \mnras, 377, 465

\bibitem[Feltzing et al.(2009)]{2009A&A...493..913F} Feltzing, S., Primas, F., \& Johnson, R.~A.\ 2009, \aap, 493, 913

\bibitem[Ferraro et al.(2018)]{2018ApJ...860...36F} Ferraro, F.~R., Lanzoni, B., Raso, S., et al.\ 2018, \apj, 860, 36

\bibitem[Gaia Collaboration et al.(2016)]{2016A&A...595A...1G} Gaia Collaboration, Prusti, T., de Bruijne, J.~H.~J., et al.\ 2016a, \aap, 595, A1

\bibitem[Gaia Collaboration et al.(2018)]{2018A&A...616A...1G} Gaia Collaboration, Brown, A.~G.~A., Vallenari, A., et al.\ 2018, \aap, 616, A1 

\bibitem[Goldsbury et al.(2010)]{2010AJ....140.1830G} Goldsbury, R., Richer, H.~B., Anderson, J., et al.\ 2010, \aj, 140, 1830-1837

\bibitem[Harris(1996)]{1996AJ....112.1487H} Harris, W.~E.\ 1996, \aj, 112, 1487

\bibitem[Heyl et al.(2017)]{2017ApJ...850..186H} Heyl, J., Caiazzo, I., Richer, H., et al.\ 2017, \apj, 850, 186

\bibitem[Hong et al.(2016)]{2016MNRAS.457.4507H} Hong, J., Vesperini, E., Sollima, A., et al.\ 2016, \mnras, 457, 4507

\bibitem[Hong et al.(2018)]{2018MNRAS.tmp.3147H} Hong, J., Patel, S., Vesperini, E., Webb, J.~J., \& Dalessandro, E.\ 2018, \mnras, arXiv:1812.01229

\bibitem[Libralato et al.(2018)]{2018ApJ...861...99L} Libralato, M., Bellini, A., van der Marel, R.~P., et al.\ 2018, \apj, 861, 99 

\bibitem[Milone et al.(2012)]{2012A&A...540A..16M} Milone, A.~P., Piotto, G., Bedin, L.~R., et al.\ 2012, \aap, 540, A16

\bibitem[Milone et al.(2017)]{2017MNRAS.464.3636M} Milone, A.~P., Piotto, G., Renzini, A., et al.\ 2017, \mnras, 464, 3636

\bibitem[Milone et al.(2018)]{2018MNRAS.479.5005M} Milone, A.~P., Marino, A.~F., Mastrobuono-Battisti, A., \& Lagioia, E.~P.\ 2018, \mnras, 479, 5005

\bibitem[Nardiello et al.(2015)]{2015MNRAS.451..312N} Nardiello, D., Piotto, G., Milone, A.~P., et al.\ 2015, \mnras, 451, 312

\bibitem[Nardiello et al.(2018)]{2018MNRAS.481.3382N} Nardiello, D., Libralato, M., Piotto, G., et al.\ 2018, \mnras, 481, 3382

\bibitem[Piotto et al.(2015)]{2015AJ....149...91P} Piotto, G., Milone, A.~P., Bedin, L.~R., et al.\ 2015, \aj, 149, 91

\bibitem[Richer et al.(2013)]{2013ApJ...771L..15R} Richer, H.~B., Heyl, J., Anderson, J., et al.\ 2013, \apjl, 771, L15

\bibitem[Tiongco et al.(2016)]{2016MNRAS.455.3693T} Tiongco, M.~A., Vesperini, E., \& Varri, A.~L.\ 2016, \mnras, 455, 3693

\bibitem[Tiongco et al.(2017)]{2017MNRAS.469..683T} Tiongco, M.~A., Vesperini, E., \& Varri, A.~L.\ 2017, \mnras, 469, 683

\bibitem[Trenti \& van der Marel(2013)]{2013MNRAS.435.3272T} Trenti, M., \& van der Marel, R.\ 2013, \mnras, 435, 3272

\bibitem[van der Marel \& Franx(1993)]{1993ApJ...407..525V} van der Marel, R.~P., \& Franx, M.\ 1993, \apj, 407, 525

\bibitem[van der Marel \& Anderson(2010)]{2010ApJ...710.1063V} van der Marel, R.~P., \& Anderson, J.\ 2010, \apj, 710, 1063
  
\bibitem[Vesperini et al.(2013)]{2013MNRAS.429.1913V} Vesperini, E., McMillan, S.~L.~W., D'Antona, F., \& D'Ercole, A.\ 2013, \mnras, 429, 1913

\bibitem[Wagner-Kaiser et al.(2017)]{2017MNRAS.468.1038W} Wagner-Kaiser, R., Sarajedini, A., von Hippel, T., et al.\ 2017, \mnras, 468, 1038

\bibitem[Webb \& Vesperini(2017)]{2017MNRAS.464.1977W} Webb, J.~J., \& Vesperini, E.\ 2017, \mnras, 464, 1977

\end{thebibliography}

\end{document}